\documentclass{sigchi}

\toappear{\scriptsize Permission to make digital or hard copies of all or part of this work for personal or classroom use is granted without fee provided that copies are not made or distributed for profit or commercial advantage and that copies bear this notice and the full citation on the first page. Copyrights for components of this work owned by others than ACM must be honored. Abstracting with credit is permitted. To copy otherwise, or republish, to post on servers or to redistribute to lists, requires prior specific permission and/or a fee. Request permissions from permissions@acm.org. \\

This is the pre-print version. The paper is published in the proceedings of the ISS 2018 conference. Final version DOI: https://doi.org/10.1145/3279778.3279802.
}

\usepackage[pdflang={en-US},pdftex]{hyperref}
\usepackage{balance}       
\usepackage{graphics}      
\usepackage[T1]{fontenc}   
\usepackage{txfonts}
\usepackage{mathptmx}
\usepackage{color}
\usepackage{booktabs}
\usepackage{textcomp}

\def\plaintitle{LeviCursor: Dexterous Interaction with a Levitating Object}

\def\emptyauthor{}
\def\plainkeywords{Ultrasonic levitation; user interaction; pointing devices; multimodal displays.}

\makeatletter
\def\url@leostyle{%
	\@ifundefined{selectfont}{
		\def\UrlFont{\sf}
	}{
		\def\UrlFont{\small\bf\ttfamily}
}}
\makeatother
\def\pprw{8.5in}
\def\pprh{11in}

\definecolor{linkColor}{RGB}{6,125,233}
\hypersetup{%
pdftitle={\plaintitle},
pdfauthor={\emptyauthor},
pdfkeywords={\plainkeywords},
pdfdisplaydoctitle=true, 
bookmarksnumbered,
pdfstartview={FitH},
colorlinks,
citecolor=black,
filecolor=black,
linkcolor=black,
urlcolor=linkColor,
breaklinks=true,
hypertexnames=false
}

\begin{document}

\title{\plaintitle}

\numberofauthors{3}
\author{%
  \alignauthor{Myroslav Bachynskyi\\
    \email{myroslav.bachynskyi@uni-bayreuth.de}}\\
  \alignauthor{Viktorija Paneva\\
    \email{viktorija.paneva@uni-bayreuth.de}}\\
    \affaddr{Department of Computer Science}\\
    \affaddr{University of Bayreuth}\\
  \alignauthor{J\"org M\"uller\\
    \email{joerg.mueller@uni-bayreuth.de}}\\
}

\teaser{
		\centering
		\includegraphics[width=\linewidth]{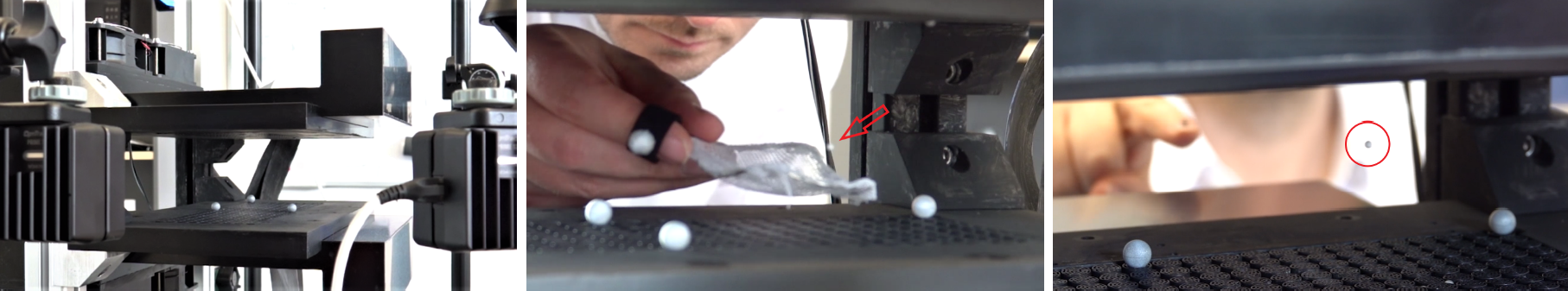}
		\caption{LeviCursor enables dexterous interactive control of levitated particles. Users can control particle motion using an optical marker attached to a fingernail. Because we use the optimization-based approach to ultrasonic levitation, particle motion is smooth in any direction. We achieve round-trip latencies of 15 ms, sub-millimeter accuracy, and stability in levitation. Please see the accompanying video for demonstration.}
	\label{fig:teaser}
}

\maketitle
%


\begin{abstract}
	We present LeviCursor, a method for interactively moving a physical, levitating particle in 3D with high agility.
	The levitating object can move continuously and smoothly in any direction.
	We optimize the transducer phases for each possible levitation point independently.
	Using precomputation, our system can determine the optimal transducer phases within a few microseconds and achieves round-trip latencies of 15 ms.
	Due to our interpolation scheme, the levitated object can be controlled almost instantaneously with sub-millimeter accuracy.
	We present a particle stabilization mechanism which ensures the levitating particle is always in the main levitation trap. 
	Lastly, we conduct the first Fitts' law-type pointing study with a real 3D cursor, where participants control the movement of the levitated cursor between two physical targets.
	The results of the user study demonstrate that using LeviCursor, users reach performance comparable to that of a mouse pointer.
\end{abstract}

{}{}{}

%
%



\section{Introduction}
One of the longest-standing visions in Human-Computer Interaction is that of the ``Ultimate Display''~\cite{sutherland}.
This entails a room in which the computer can control the existence of matter.
The computer could create chairs, or bullets in such a room, and the virtual and physical worlds would truly be merged.

One approach to creating an ``Ultimate Display'' has resulted in Programmable Matter~\cite{goldstein} and Radical Atoms~\cite{ishii}.
Programmable Matter would consist of millions of miniature robots. 
The main difficulties associated with this ``active atom'' approach are ensuring sufficient and reliable power supply to the individual units, costs per unit, and miniaturization of the units.

To overcome the limitations of the ``active atom'' approach, an alternative is to use ``passive atoms''.
In this approach, the ``atoms'' themselves are passive, but actuation, power supply and intelligence are provided by the environment.
This solves the three problems referred to above.
This concept was for example employed by Pixie Dust~\cite{pixiedust}.

Pixie Dust uses phased arrays of ultrasonic transducers to generate acoustic standing waves and create a grid of nodes, where small objects can be levitated.

From a Human-Computer Interaction perspective, an important problem of the ``Ultimate Display'' approach is how to interact, for example, how to move particles interactively.
Pixie Dust~\cite{pixiedust} explores interactive particle control methods, but the use of the classical standing-wave approach for trap generation, introduces limitation in the movement of the particle. 
While the traps are quite stable, smooth movement is only possible in one dimension between two opposing arrays or array and reflector. 
Thus smooth movement in 3D would require six opposing arrays. 
This would significantly impair the visibility of the levitating display. 
The Pixie Dust setup consists of four transducer array, allowing for smooth particle movement in 2D. 
In addition, levitation is only possible within the boundaries of the arrays and is limited to parallel array arrangements. 

LeviPath~\cite{levipath} provides an algorithm for moving levitated particles with two opposing arrays, on a 3D grid. 
The phase values at each step, of approximately $0.2$~mm, are precomputed and stored in a table. 
However, as shown in the video \cite{levipath:video}, the particles move at a relatively low speed and experience some jittering. 

JOLED~\cite{joled} uses optimization for phase computation, which enables smoother particle movement than the pure standing-wave approach. 
The JOLED setup is composed of 60 transducers in total.
Due to the low number of transducers, real-time particle control using mouse or keyboard is possible. As a consequence, however, the display volume is relatively small.

In general, particle interaction has been implemented at relatively low speeds, due to the risk of particles being dropped during high speeds or high accelerations.
In summary, although interaction with levitating particles has been explored by Ochiai et al.~\cite{pixiedust}, Omirou et al.~\cite{levipath} and Sahoo et al.~\cite{joled}, real time gesture interaction with a particle moving along a smooth 3D path with high speed has not been yet achieved.


In this paper, we address the problem of dexterous interactive movement of levitated particles.
The main difficulties in achieving this with homogeneous movement along all three dimensions are ensuring:
(1) low latency,
(2) continuous movement without steps, and
(3) stable movement enabling high velocities and accelerations. 

We use the optimization-based approach of Marzo et al.~\cite{marzo:nature}, avoiding the inhomogeneties of the classical standing-wave levitation. 
The main limitation of applying the optimization approach in~\cite{marzo:nature} to our setup is that, due to the larger number of transducers, optimization takes about $1$~s for each levitation point, thus preventing interactive rates.
We solve this problem and achieve (1) low latency by precomputing optimal phases for \textit{all possible levitation points within the entire array at 0.5~mm resolution}, resulting in a round-trip latency of 15~ms.
Jumps of the trap, even if just by 0.5~mm, result in noticable jumps followed by oscillations of the particle. 
We achieve (2) continuous movement by interpolating between the precomputed levitation points at 1~kHz, achieving arbitrarily small step sizes.
Optimization creates numerous weaker traps in the vicinity of the main trap.
Previously, when placing the particle, one could not be sure that it is actually located in the main trap.
Furthermore, over time, the particle might jump to weaker secondary traps, resulting in offset and reduced stability. 
We achieve (3) stable movement by providing a mechanism to stabilize levitation and ensure the particle is always in the main trap.

The LeviCursor method can be beneficial for studies and applications involving 3D selection with a physical object as cursor, where the correct perception of the 3D targets and the 3D cursor is crucial. 
It provides a novel method of interacting with tangible interfaces, while opening up new research questions in the HCI community concerning perception, motor control and transfer function of physical cursors which are detached from the user's body. 
In addition to pointing and selection, precise and accurate manipulation of levitating particles can be used to improve graphical visualizations and animations in mid-air~\cite{pixiedust}, provide better gaming experience in levitation-based games~\cite{joled} as well as facilitate containerless handling and mixture of sensitive materials, i.e. lab-in-a-drop~\cite{drops}, in favor of preventing contamination.

\section{Related work}
\subsection{Acoustic Levitation}
%
%
Acoustic radiation force can be used to counteract gravity and trap millimeter-sized objects in mid-air.
This effect is most often achieved by using phased arrays of ultrasonic sound emitters of the appropriate phase and amplitude to create acoustic nodes in mid-air, where particles can be trapped.  
%
Acoustic levitation does not require any special (e.g. optical, magnetic, electric etc.) properties of the levitating object.
Therefore a variety of objects can be levitated, including solids, liquids and insects~\cite{tinylev}.
Furthermore, particles of smaller (i.e. Rayleigh particles)~\cite{pixiedust} and larger (i.e. Mie particles)~\cite{mie} radius than the wavelength have been levitated. 

\subsection{Moving Levitated Particles}
A few methods for achieving controlled movement of levitating particles in the acoustic field have already been developed.
\textit{LeviPath}~\cite{levipath} employs an algorithm which combines basic patterns of movement to levitate objects across 3D paths, in a setup consisting of two opposed arrays of transducers.
The input path is decomposed into a height variation, controlled by the phase difference between the top and bottom transducer array and a 2D path.
The 2D path is then adapted to a possible pattern, obtained by interpolation between adjacent pairs of  levitating points.
In addition to controlled translational movement in the field, controlled rotations have also been achieved, but with the help of electrostatic forces.  In \textit{JOLED}~\cite{joled} levitating particles of different physical properties are coated with titanium dioxide in order to induce electrostatic charge.
This allows for the control of the angular position of the particles by the means of electrostatic rotation. The 3D position of the particles is determined by optimizing the phases of the acoustic arrays.

\subsection{Interaction with Levitated Particles}
For the purpose of contactless manipulation of particles using acoustic levitation, the wearable glove \textit{GauntLev}~\cite{gauntlev}, with integrated ultrasonic transducers, has been designed. The
\textit{GauntLev} gloves trap particles either in front of the palm or between a pair of fingers, enabling a set of basic maneuvers such as capturing, transferring and combining levitating particles, a process performed manually or computer assisted. Alternative devices that can be used to manipulate levitated particles which are not attached to the hand are the \textit{Sonic Screwdriver}, a parabolic head with a handle that can generate twin traps and \textit{UltraTongs}, tweezers that generate standing waves~\cite{gauntlev}. 
Some of the configurations in~\cite{gauntlev} and~\cite{marzo:nature} support one-sided levitation, which provides very good display visibility, but achieving fast and stable levitation is more challenging. 

Concerning levitation with static acoustic elements, thus far, only interaction techniques for selection and step translation of particles have been developed.
With \textit{Point-and-Shake}~\cite{pointandshake}, users can point a finger to select levitating objects and receive visual feedback in the form of a continuous side-to-side (\textit{shake}) movement.
The hand gestures are tracked using a Leap Motion sensor.
Interactive ontrol of a single levitated particle using a keyboard, mouse, GUI buttons and a Leap Motion sensor was presented in \textit{LeviPath}~\cite{levipath}. The particle was moved by small steps on a 3D grid. 
The \textit{Pixie Dust}~\cite{pixiedust} setup comprises four vertical transducer arrays, facing inwards, which generate a 2D grid of acoustic nodes.
Interactive techniques were tested either by using Kinect to detect users' hand gestures, which were then mapped to a particular particle path in the acoustic field (e.g. translating a cluster of particles along one horizontal axis) or by using a pointing touch screen device to assign the trajectories.

\subsection{Summary}
A variety of approaches to ultrasonic levitation have been developed.
However, dexterous interaction with levitated objects has not yet been demonstrated.
For approaches using only standing waves (in the form of focal lines), the main limitation is that different techniques must be used to move particles in different dimensions.
Up to now, this has resulted in less smooth, less agile and often jumpy object motion.
Marzo's \cite{marzo:nature} optimization approach allows for continuous placement of traps at arbitrary locations within the working volume.
By displacing these traps with small amounts (approx. 0.1 mm), continuous particle motion can be achieved.
In \cite{marzo:nature} real-time interaction with the system was possible using a keyboard or GUI buttons, however the rates are still too slow for continuous interaction.
On larger setups, the optimization would take several seconds for each location.
Up to now, this prevented a smooth interactive use of this technique.

Our paper contributes the first implementation of a low-latency, high frame-rate, smooth interactive control of a levitated particle in 3D space, as well as a method that ensures sustained particle positioning in the main trap. 
In addition, we conduct the first device-mediated Fitts' law study in 3D with a levitated particle as cursor, providing all natural depth cues.


\section{System}
Our main challenges are to achieve homogeneous movement along all three dimensions with:
(1) low latency,
(2) continuous movement without steps, and
(3) stable movement enabling high velocities and accelerations.

We overcome these challenges by:
(1) Precomputation of optimal transducer phases for \textit{all possible levitation points within the entire array at 0.5~mm resolution}.
(2) Phase interpolation.
(3) A particle stabilization mechanism to ensure that the particle is always in the main trap.

\subsection{Precomputation of Optimal Transducer Phases}
The main limitation of using the optimization-based approach to render interactive levitating interfaces, is that optimization can take several seconds for each new point.
We update the levitation points at $1$~kHz, rendering this approach to be unfeasible.
\cite{marzo:nature} presents an approach to precomputing discrete animation paths which can then be played back.
We extend this approach to precompute {\em all levitation points in the entire levitation volume at $0.5$~mm discretization}.
Our levitation volume measures $140$~mm width * $80$~mm height * $90$~mm depth.
At $0.5$~mm resolution, this results in approximately 8 mio. levitation points.
For each of these points, the 252 transducer phases have to be optimized. 
We optimize each point using $20000$ iterations of BFGS.
We use Armijo line search with coefficient $\alpha=0.8$ to determine the step size.
This takes about 20 seconds per point.
The entire calculation takes approx. 44800 hours (> 5 years) of computation time.
Since calculation on a workstation is not feasible, we resort to using a computer cluster.
We stored the result in a lookup table with a size of $8$ GB in RAM.

\begin{figure}[t]
	\centering
	\includegraphics[width=1\columnwidth]{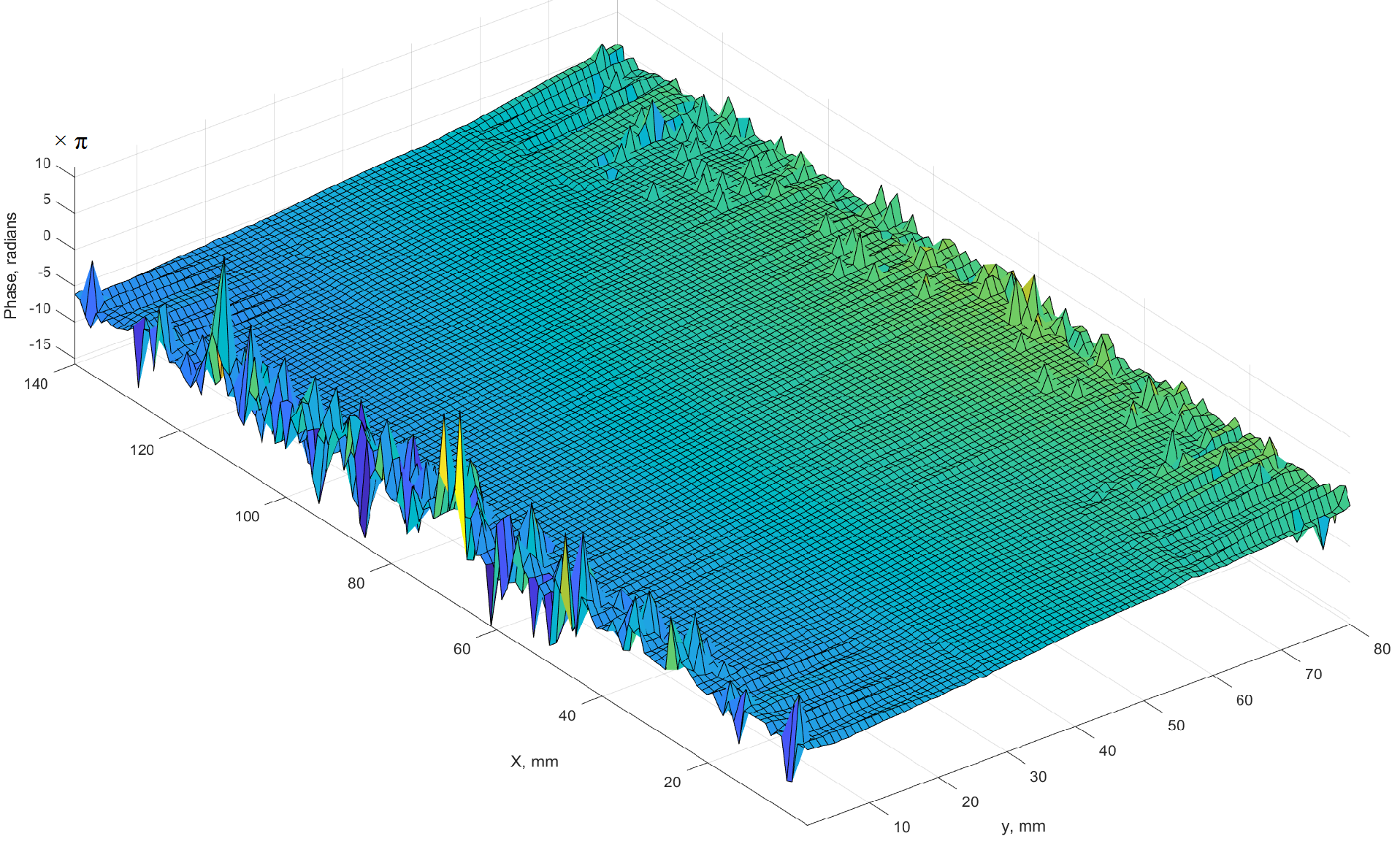}
	\caption{Evaluation of phase smoothness within the sound-field volume. The surface represents phases of the transducer 197 at the plane $Z=9$~mm.}
	\label{fig:smoothness}
\end{figure}

Because we interpolate the phases between levitation points, it is very important that the phases for neighboring points are smooth.
Unfortunately, the optimization problem inherently contains many local optima. 
Ideally, neighboring points should use the ``same'' local optimum, and avoid jumping to a distant one, as such a transition would render the interpolated data inconsistent and lead to unpredictable behavior of the levitated particle.
After evaluating diverse approaches to achieving this, we propose the following strategy.
First, the center of the levitation volume is optimized from random starting phases.
Any subsequent point is optimized using the phase values of a neighboring point for starting phases.
After the center point, we optimize progressively in the height dimension (up and down). 
To ensure smoothness, we optimize with 0.1 mm resolution.
From this line, we optimize the entire width of the array with 0.1 mm resolution.
This results in an optimized plane at depth 0.
From this plane, we optimize in the depth dimension at 0.5 mm resolution. 
This procedure results in very smooth transducer phases between neighboring points (see Figure~\ref{fig:smoothness}). 
Any remaining non-smoothness is mostly in the height dimension. 

\subsection{Phase Interpolation}
In order to achieve sub-millimeter precision in manipulation of the sound field, we use trilinear interpolation between the eight neighboring points from the lookup table.
We first evaluate the acceptability of such interpolation by numerically computing smoothness within the whole sound-field volume.
We consider the transition between two neighboring points as smooth, if the differences between the phase values of each transducer are not larger than $\pi$ radians.
The majority (96.2\%) of the phase transitions within the sound field volume are smooth and far smaller than $\pi$.
However, there is still a small fraction of non-smooth transitions, which needs to be investigated.
We inspect spatial properties of the transition smoothness, and in particular those of non-smooth transitions, using visualizations of the transducer phases over multiple slice surfaces within the volume (Figure~\ref{fig:smoothness}).
As can be observed, the phases are smooth close to the center of the volume, and become non-smooth closer to the boundaries, in particular in the proximity of the transducers.
Based on our observations, we configured the trilinear interpolation so that it is applied if the neighborhood of the point is smooth, and it is replaced by the nearest neighbor values if the neighborhood is non-smooth.
The particle movement is less smooth (0.5~mm steps) when entering a non-smooth region, but the general stability of the particle movement is increased.

\subsection{Particle Stabilization}
One major problem with ultrasonic levitation is placing the particles. 
When a focus point is generated by the optimizer, weaker secondary traps also appear in the acoustic field.
These secondary traps can levitate particles, but are prone to disappear and drop them once the primary trap is moved.
Since the acoustic field cannot be seen with the naked eye, one can not distinguish between different traps. 
Consequently, placing the particle in the main trap is not a trivial task.
Furthermore, after some time, the particle may jump to a secondary trap.

\begin{figure}[b]
	\centering
	\includegraphics[width=\columnwidth]{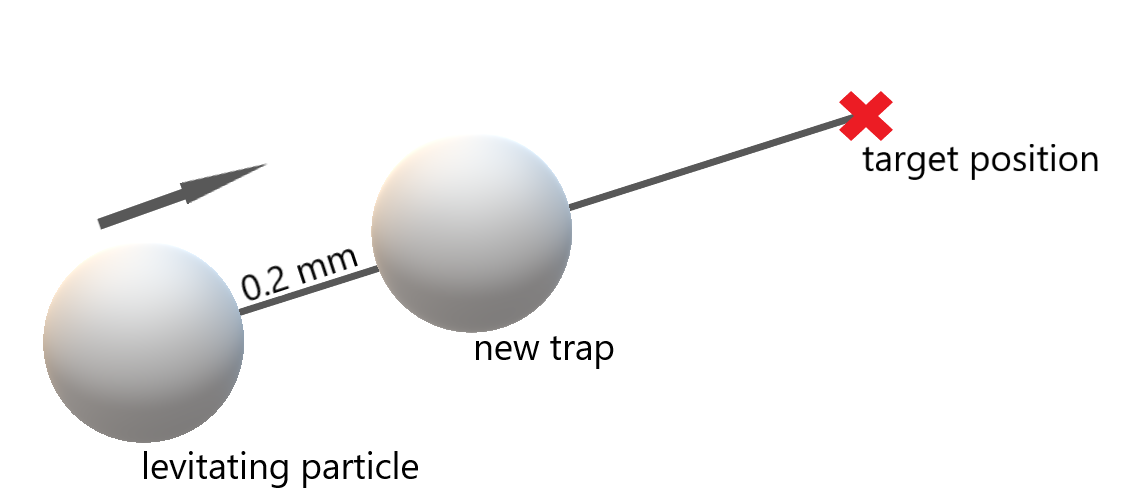}
	\caption{When the particle is moving towards a new target position, it never takes steps larger than 0.2 mm per frame, to ensure that it stays in the primary trap.}
	\label{fig:stabilization}
\end{figure}

We stabilize the particle, i.e. reassure it is located in the primary trap,  both when the particle is first placed into the acoustic field as well as during direct interaction. 
When placing the particle, we optimize the field for a levitation point at the origin. 
We place the particle in the acoustic field, using a piece of an acoustically-transparent fabric.
Then we turn on the transducers, which causes levitation of the particle in some secondary trap.
We determine the actual particle position using the motion capture system and generate a primary trap at the actual particle position.
During interactive control of the particle, excessively large jumps of the primary trap can cause the particle to jump into a secondary trap.
Therefore, we interpolate the primary trap position towards the target indicated by the user, while ensuring that the primary trap never moves more than 0.2~mm between frames in the regions with interpolation.
In Figure~\ref{fig:stabilization}, a levitating particle moving towards a new target position is shown. 
In the subsequent frame, a new primary trap is generated in the direction of the target, at a distance of 0.2~mm.
This procedure contributes substantially to the stability of the levitated particle. 

\section{Hardware}
Our acoustic levitator
comprises two $9\times 14$ arrays of \textit{muRata MA40S4S} transducers.
The transducers are cylindrical and have a $10$~mm diameter and a $7$~mm height.
The ultrasonic transducers are equally spaced at a distance of $0.3$~mm from each other and have maximum input voltage of $20$~Vpp.
Each emits a sound wave of frequency $f=40$~kHz (wavelength $\lambda = 8.6$~mm), which is inaudible to humans.
The two arrays are mounted horizontally, facing each other, at a distance of $80$~mm.
We developed an aluminum rail system, which allows for easy adjustment of the distance between the arrays.

A major problem when using transducer arrays for levitation is that the arrays heat up fast, leading to destruction within a few minutes.
We solved this problem with a cooling system that generates an air stream on the back of the array PCBs, without leaking an air stream into the levitation volume.
This allows us to operate the arrays continuously.
We use expanded polystyrene beads of small diameter (approx. 2~mm) as levitating particles, due to their low density.

For driving the transducer arrays, we use the logic board of the \textit{Ultrahaptics} \footnote{http://ultrahaptics.com} Evaluation Kit.
We connected the board to both transducer arrays, leading to on-board synchronization of both arrays.
The logic board is connected to a driving PC using USB. 

We track the particle position and index finger of the user using optical motion capture (\textit{OptiTrack}). 
We use a small velcro-attached retro-reflective marker with a diameter of $9$~mm, placed directly on top of the user's fingertip.
We use six \textit{Prime 13} infrared cameras capturing 240 FPS.
Three cameras observe the levitation volume from the side, while three additional cameras track the user's finger from above.
The cameras are connected via Ethernet to a second PC that drives the motion capture system and our levitation software.

\section{Software}
Our precomputation software is based on the system implemented by Marzo et al, which is generously shared in~\cite{asier:ultraino}.
Based on this, we developed a program for phase optimization that is suited for execution on a computer cluster.
We slice the workload into 88000 task description files using a script.
Worker nodes read these files and generate a result file.

The interactive hardware and software has to operate in real-time. 
We use two workstations to operate the system, so as to reduce latencies. 
The first workstation operates in high-performance mode and runs the \textit{OptiTrack Motive} motion-capture system. 
Particle and fingertip are tracked and streamed via NatNet to a custom Java program running on the same machine.
The Java program performs particle stabilization and computes the particle motion.
This program reads in the results files from the cluster computation at startup, so as to generate the lookup table.
It looks up the necessary transducer phases in this table.
Finally, it performs phase interpolation and sends the resulting transducer phases to a C++ program on a second workstation.

The second workstation is tuned to run the C++ application which receives the transducers' states through a UDP socket. 
The C++ program caches the phases locally and uses the Ultrahaptics Low-level SDK to stream the phases to the Ultrahaptics logic board. 
To ensure smooth levitation, the C++ software needs to respond to a callback from the Ultrahaptics driver at 1~kHz with a latency of a few milliseconds at maximum. 
This workstation runs only the critical operating system processes with low priority on one half of the CPU cores, as defined by an affinity mask.
The real-time priority and the other half of the cores (non-hyperthreaded) are dedicated to the C++ application.
The machine runs in high-performance mode with CPU sleep states and SpeedStep disabled.
Both workstations are connected via Ethernet using a local Gigabit switch.
The experiment is controlled and logged using the Java program on the first workstation, which also computes particle motion. 

\section{Technical evaluation}

To evaluate velocities and stability, similarly to LeviPath~\cite{levipath}, we performed an experiment in which we moved a particle back and forth within the levitation volume along a $7$~cm straight path. 
We repeated the movement five times at each velocity and recorded the number of successes and failures. 
When the particle correctly completed the full movement along the given path, success was registered. 
A failure was noted when the particle fell off or switched to a secondary trap during the movement.
We started with a velocity of $0.2$~m/s, gradually increasing it by steps of $0.2$~m/s up to $1.2$~m/s, where failure was observed in all five trials.

\begin{figure}[t]
	\includegraphics[width=\linewidth]{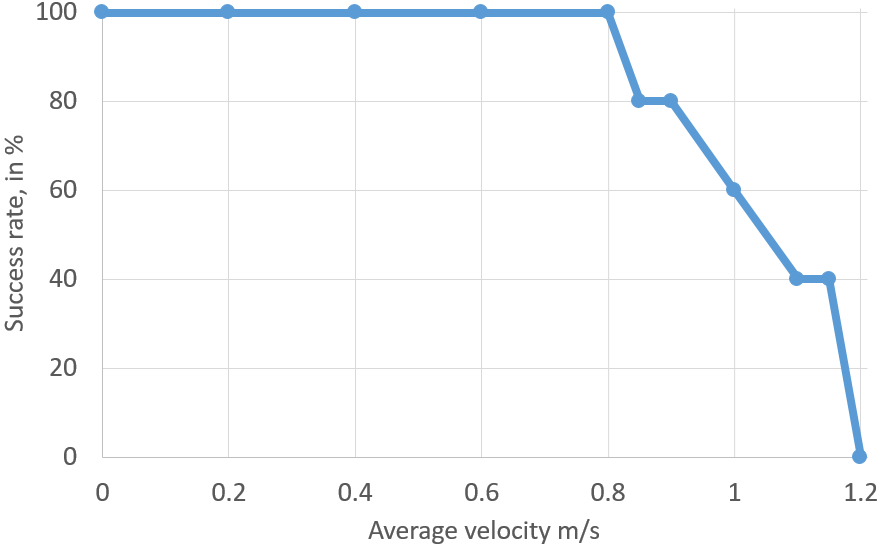}
	\caption{Success rate with respect to average particle velocity.}
	\label{fig:stability}
\end{figure}

As can be seen in Figure~\ref{fig:stability}, our system achieved particle velocity of $0.8$~m/s with a $100\%$ success rate, thereafter the success rate decreased almost linearly and eventually reached $0$ at a velocity of $1.2$~m/s.
From this experiment we can conclude a lower bound on the maximum velocity of $0.8$~m/s. 
We observed, however, that most of the failure cases consisted of the particle dropping either at the beginning or at the end of the movement.
This indicates that the limiting factor is not the velocity, but the acceleration. 
In fact, we believe that by providing more dynamically consistent control it should be possible to achieve even higher particle velocities.
For example, our system was able to achieve velocities close to $1.5$~m/s, however in this case the particle was shooting out of the end-trap.
In the future, we want to conduct experiments where the maximum reachable velocity and acceleration in the right-most part of Figure~\ref{fig:stability} ($0.8$ to $1.2$~m/s), are explored separately.


We also evaluated the total latency of the system using a high frame-rate camera.
We setup a motion capture marker-based event (marker crossing a plane) and a response of the levitation system (dropping the currently levitated particle).
The camera observed the space where both event and response were generated and recorded the corresponding segments.
We repeated the experiment three times and tallied the number of frames between the marker event and the system response.
For a system to be perceived as real-time in pointing tasks, the total latency has to be below $20$~ms~\cite{jota2013}.
In our experiment, in all three cases the latency between the event and the response was less than $17$~ms, with the average value being $15$~ms, which is below the threshold value perceptible for users.

\section{User Study}
As suggested in the introduction, key application areas allowed by LeviCursor are physical 3D pointing, including 3D pointing with tangibles, and aimed movement user studies providing all natural depth cues of the cursor and the targets.
LeviCursor allows user studies of mediated 3D pointing to investigate effects of latency, control-to-display ratio or the transfer function on the pointing process, accuracy, speed, physical ergonomics, cognitive load, movement dynamics, velocity and acceleration profiles etc.
There are multiple user studies investigating pointing movements in 3D space, however in contrast to LeviCursor they provide either limited cues for depth perception e.g. using volumetric display~\cite{grossman2004} or virtual reality~\cite{teather2011}, or they do not allow for any transfer function, for example non-mediated 3D pointing~\cite{Bachynskyi:2015:IDN:2722827.2687921}.
We demonstrate applicability of LeviCursor to pointing tasks by running a short user study of 3D aimed movements.

\begin{figure}[h]
	\begin{minipage}{\columnwidth}
		\centering
		\includegraphics[width=0.9\linewidth]{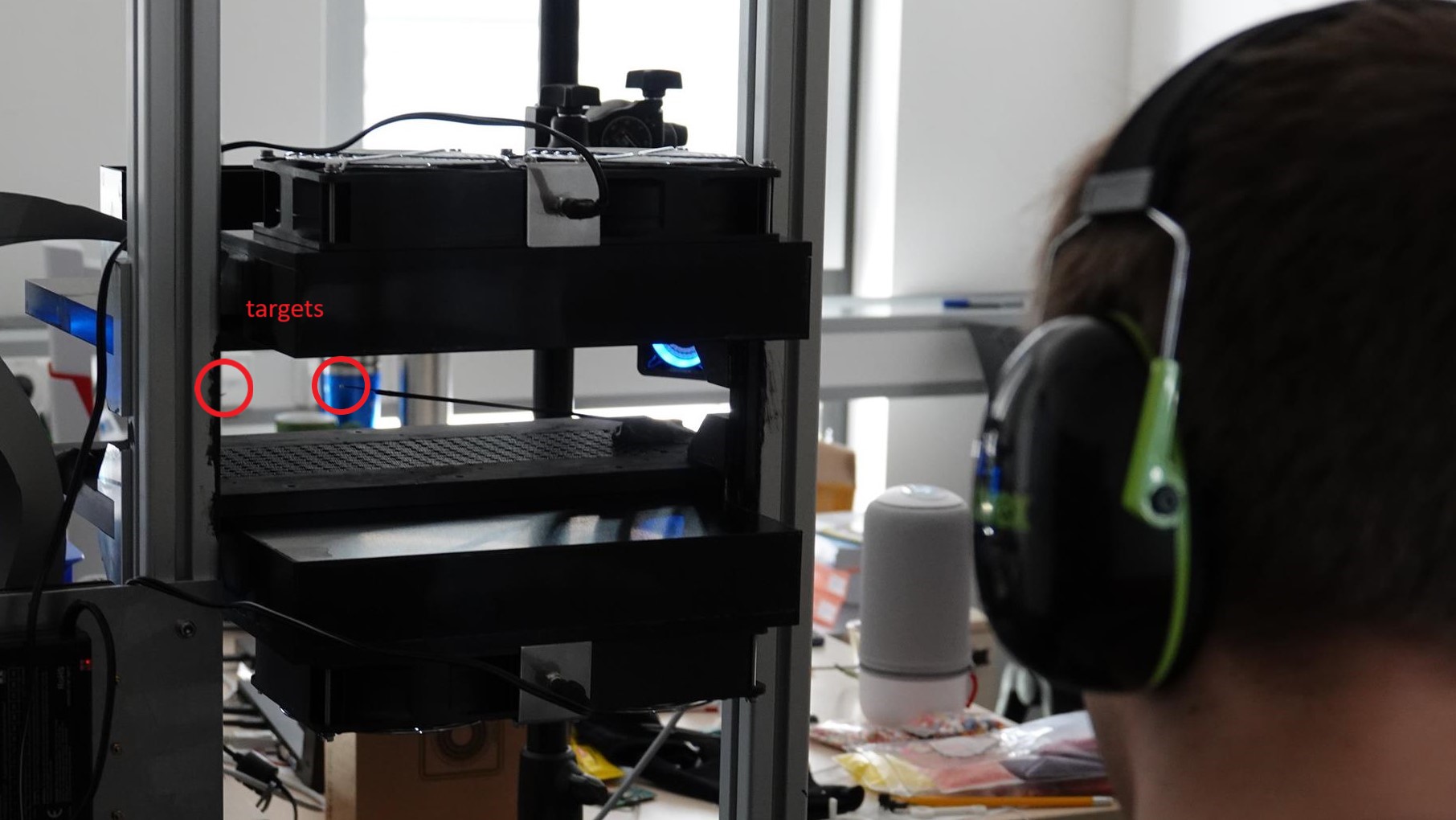}
	\end{minipage}
	\begin{minipage}{\columnwidth}
		\centering
		\includegraphics[width=0.9\linewidth]{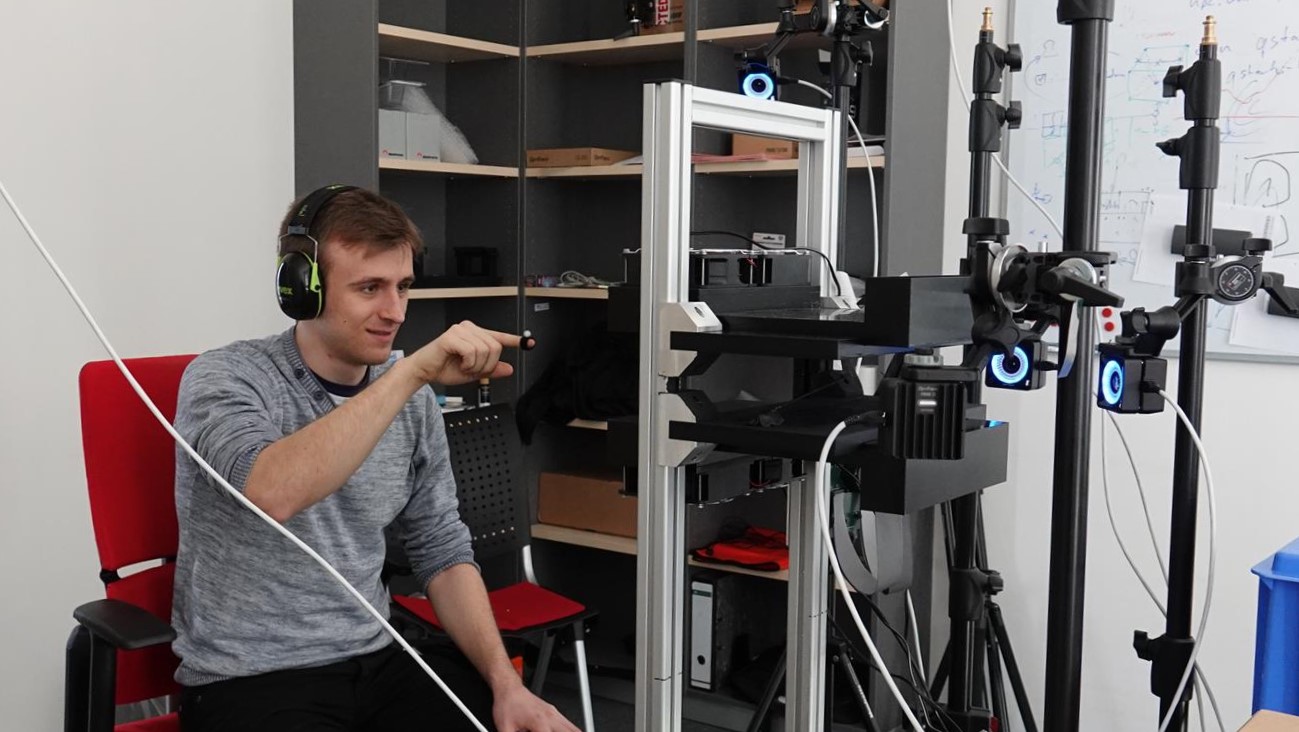}
	\end{minipage}
	\caption{Participants had a retroreflective marker attached to their right index finger and were seated on a chair in front of the levitation apparatus. With their fingertip, they were able to control the levitating particle in front of them and complete the given pointing tasks. The two targets within the levitation volume are marked with red.}
	\label{fig:user_study}
\end{figure}

The task was a variation of Fitts' {\it serial pointing task} adapted to 3D. 
It is very difficult to place physical targets for levitating particles.
The targets should disturb physical particle motion, the sound field, and the motion capture system as little as possible.
We decided to use needles painted with black matte color to show the center of the targets.
The actual targets were internally represented as spheres around the needle tips and were registered using the motion capture system.
We used three target sizes of: 2~mm, 4~mm and 8~mm radius. 
The distance between the targets was 68 mm. 
The target size conditions for each user were randomized.
The task of the user was to move the particle between the two targets as quickly as possible. 
The motion capture system was tracking the position of the particle with respect to both targets.
When the particle entered the target, a confirmation tone sounded and a success was registered.

We recruited 8 participants (mean age 30.5 years, std. dev. 5.6, 4 male, all normal or corrected to normal eyesight, all right-handed).
Participants sat on a chair in front of the apparatus (see Figure~\ref{fig:user_study}).
A retroreflective marker of 9 mm diameter was attached to the index finger of their right hand.
The particle was placed in the levitation volume by the experimenter.
Participants could control particle motion in 3D with their fingertip, using a control-to-display ratio of 3.
Participants were allowed to explore the particle motion for approx. 30~s.
We asked participants to place the particle as accurately at each of the needle tips as they could, in order to calibrate the target location according to their perceptions of the target.
Afterwards, participants were asked to move between the targets as quickly as possible.
After performing 50 aimed movements, the experiment was shifted to the next target size-condition.

During the experiment, our software was continuously recording the 3D position of the particle, the real-time timestamps and the timestamps when the user reached each target and was notified by the sound.
After the experiment, the participants were informally interviewed concerning their experience with LeviCursor.

\subsection{Analysis}
We applied Fitts' law analysis, as is typical for the HCI field~\cite{mackenzie1992}.
While there exist multivariate models of pointing~\cite{grossman2004}, for spherical targets they are equivalent to Fitts' law.
We use Fitts' law in the Shannon formulation
\begin{equation*}
MT=a+b\times\log_2\left(\frac{D}{W}+1\right),
\end{equation*}
where $MT$ is the movement time, $D$ the amplitude, $W$ the target width, $a$ and $b$ are free regression coefficients. 
Following the recommendations of~\cite{mackenzie1992}, instead of $D$ and $W$, we use effective target width $W_e$, based on the standard deviation of the end-points ($\sigma$) as \[W_e=4.133\sigma\]
and the effective amplitude $D_e$ as the distance between the corresponding effective target centroids:
\[D_e=\sum_{i=1}^{N}\frac{D_i}{N},\]
where $D_i$ is the amplitude of individual aimed movement and $N$ is the number of movements terminating within the effective target.
We group the data into six ranges according to the ID.
We average IDs and MTs within each group and then fit a Fitts' law model as a first-degree polynomial optimally representing the data in the least-squares sense.
We evaluate goodness-of-fit using the coefficient of determination ($R^2$).
To evaluate the performance of the users using LeviCursor, we compute the average effective throughput
\[TP_{e a}=\frac{1}{P}\sum_{i=1}^{P}\left(\frac{1}{C}\sum_{j=1}^{C}\frac{ID_{e ij}}{MT_{ij}}\right),\]
where $P$ is the number of participants and $C$ is the number of conditions, as well as maximum effective throughput
\[TP_{e max}=\max_{i=1}^{P}\left(\max_{j=1}^{C}\frac{ID_{e ij}}{MT_{ij}}\right).\]

\subsection{Results}
\begin{figure}[b]
	\centering
	\includegraphics[width=\columnwidth]{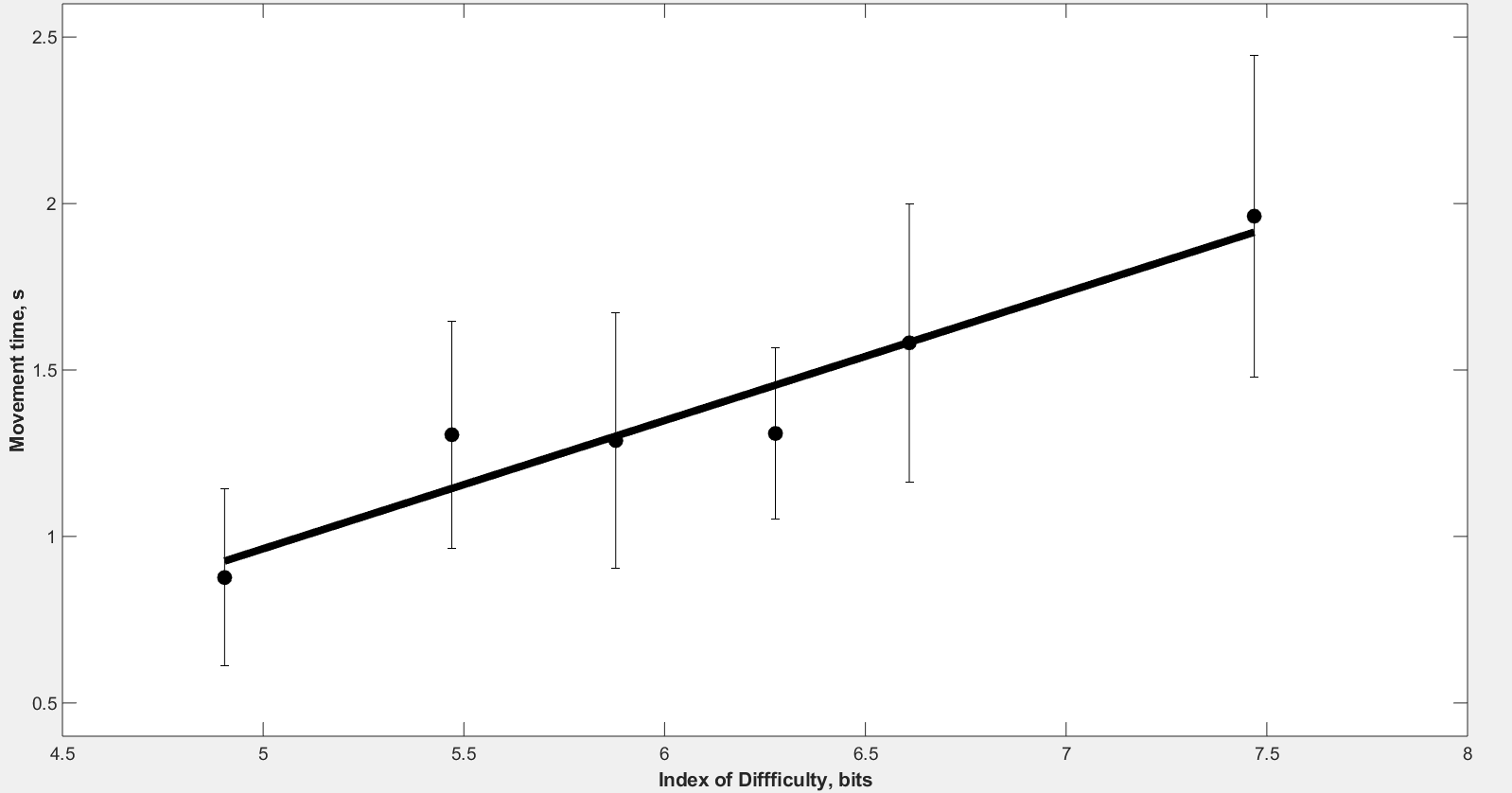}
	\caption{Fitts' law model representing the data of all participants.}
	\label{fig:fitts}
\end{figure}

The experimental data can be modeled successfully by Fitts' law with $R^2$ of 0.92, as can be seen in Figure~\ref{fig:fitts}. 
The participants achieved an average throughput of $4.93~bits/s$ and a maximum throughput of $8.69~bits/s$.
These values are comparable to the throughput of the mouse~\cite{SOUKOREFF2004751}.
Furthermore, they are only slightly below the throughput of uninstrumented mid-air pointing (average $TP=5.48~bits/s$~\cite{Bachynskyi:2015:IDN:2722827.2687921}).

According to the informal interviews, the users experienced the interaction as exciting.
It was described, for example, as "Jedi using the force", and they felt "in control of the particle".
Some of them mentioned a common problem in mid-air interaction - tension and fatigue in the shoulder, known as the "Gorilla arm".

We find it promising that even though LeviCursor has different physical properties than a virtual mouse-controlled cursor on a desktop, it can provide comparable interaction behavior and performance.
This demonstrates that using our method, users can exercise dexterous control over levitated particles. 

This was, however, a preliminary study to test a new concept. 
We plan to conduct bigger studies with more participants in future work.

\section{Discussion}
From the results of both the technical evaluation and the user study, we can clearly see that the proposed method for interactive control of levitated particles is an effective tool for applications which require pointing in the real 3D space.
While this is the first paper which demonstrates such smooth and dexterous control of levitated particles, the method also has multiple limitations and large potential for further improvement.
Below we describe the limitations and as future work we plan to explore new approaches to workaround the main limitations.

\subsection{Limitations}
The limitations of LeviCursor can be split into two parts - first the limitations inherited from the underlying levitation algorithm~\cite{marzo:nature} and second the limitations of the current algorithm.

The inherited limitations of the method relate to the optimization approach, namely we can levitate a single particle of size smaller than half of the wave length, preferably spherical (although we have also levitated flat and ellipsoidal particles), made of low-density materials.
Levitation of multiple particles should be made possible by changing the objective function for the optimization or using a method similar to~\cite{marzo:nature}.
Although ultrasound technology has passed safety tests and is cleared for commercial use for haptic and parametric audio devices (e.g. \textit{Ultrahaptics}, \textit{Ultrasonic Audio} etc.), there are still concerns about the effects of high-intensity ultrasound on humans. 
As a cautionary measure, we provided the participants of the user study with earmuffs.

The approach described in this paper also has multiple limitations, in particular: scalability with respect to the acoustic volume and the computational power necessary for precomputation, flexibility of the ultrasound array setup, extensive hardware both for ultrasound levitation and for motion tracking, and in the current implementation with optical motion tracking - color of the particle and the surroundings.
The scalability is limited by the size of the lookup table and the necessary precomputation time. 
The required memory and computational time scale linearly in each dimension. 
While in this paper we work with a levitating interface of relatively small acoustic volume, the state of the art hardware and software can allow significantly larger setups, for example current supported size of main memory (2TB by Windows) allows a levitation volume of $1.2$~m$^3$ while keeping the entire table in RAM. 
Considering that the current lookup table is computed by a cluster within few hours, it should be possible to compute the table for the above mentioned movement volume in reasonable time.
In regard to flexibility, it is necessary to recompute the lookup table for each ultrasound array setup, which takes significant computation time.
Apart from sophisticated ultrasound hardware, the current approach also requires optical motion capture hardware.
The optical motion capture cameras need to be positioned in a way, that allows the levitated particle to be visible across the entire volume of the levitating display.
As an additional requirement, the particle has to provide high visual contrast in comparison to the surrounding hardware, in the optimal case it should be retroreflective.

\subsection{Future work}
There are multiple potential improvements to the current approach of interactive control of levitated particles, as well as extensions and additional applications.

As a main direction of our future work we plan to apply other algorithms for levitation which can work in real time instead of the lookup table, namely we plan to work on using holographic acoustic elements (focus point and signature)~\cite{marzo:nature} for computation of the levitation trap in real time. Up to this point we have tried the focus and signature approach, but it was less stable than the optimized phases from the lookup table.

Next, we would like to explore levitation with multiple particles as well as interactive control of them.

Lastly, we would like to identify and test additional realms that can benefit from the LeviCursor method. 

\section{Conclusion}
In this paper, we presented LeviCursor, a method for interactively moving a 3D physical pointer in mid-air with high agility.
The method allows a levitated particle to move continuously in any direction.
We addressed the three problems of low latency, continuous movement without steps, and stable movement enabling high velocities and accelerations.
We contribute three solutions for solving these problems. 
The first is a complete precomputation of all transducer phases, achieving a round-trip latency of 15 ms.
The second is a 3D interpolation scheme, allowing the levitated object to be controlled almost instantaneously with sub-millimeter accuracy.
Lastly, we presented a particle stabilization mechanism which ensures that the particle is always in the main levitation trap. 
This interactive system has been validated by a user study. 
The results of the study showed that interaction with LeviCursor can be successfully modeled by Fitts' law, with throughput that is comparable to interaction with mouse pointers. 

\section{Acknowledgments}
This research has received funding from the European Union's Horizon 2020 research and innovation programme under grant agreement \#737087 (Levitate).


\bibliographystyle{SIGCHI-Reference-Format}
\bibliography{sample}

\appendix  
\section{Methods}
We use the same method proposed by Marzo et al.~\cite{marzo:nature} to model acoustic levitation. 
In this section, we provide an overview of this method for the reader.

The acoustic radiation force $F$ on a small (radius $<<$ wavelength), spherical particle in an invicid medium is given by the gradient of the Gor'kov potential $U$~\cite{gorkov}~\cite{bruus}
\begin{equation} \label{eqn_force}
F = - \nabla U.
\end{equation}
The equation 
\begin{equation} \label{eqn_Gor'kov}
U = k_1(\mid{p}\mid^2)-k_2(\mid{p_x}\mid^2+\mid{p_y}\mid^2+\mid{p_z}\mid^2),
\end{equation}
describes the Gor'kov potential~\cite{marzo:nature}. 
The complex modulus denoted by $\mid \cdot \mid$ is defined as $\mid \beta+i\gamma \mid=\sqrt{\beta^2+\gamma^2}$. 
Equation (\ref{eqn_Gor'kov}) consists of two parts; the first is a complex pressure part, which shows that particles move from areas of high modulus of the complex pressure $p$, to areas of low pressure modulus and the second - a velocity part, written in terms of the spatial pressure derivatives, describing how particles are drawn to areas of large modulus of the velocity gradient.
The two constants are given by:
\begin{align*} 
	k_1 &= \frac{1}{4}V\left(\frac{1}{c_m^2p_m}-\frac{1}{c_p^2p_p} \right),  \\ 
	k_2 &= \frac{3}{4}V\left(\frac{\rho_m-\rho_p}{\omega^2\rho_m(\rho_m+2\rho_p)}\right),
\end{align*}
where $V$ is the volume of the (spherical) levitating particle and $\omega$ the wave frequency.
$c_m$ and $c_p$ denote the speed of sound through the medium and the particle respectively. 
The density of the medium is given by $\rho_m$, and of the particle by $\rho_p$. 
In our case, the medium is air and the particle is an expanded polystyrene bead, so we have: $c_m=343\frac{m}{s}$, $c_p=2400\frac{m}{s}$, $\rho_m=1.2\frac{kg}{m^3}$ and $\rho_p=25\frac{kg}{m^3}$ at a room temperature of $20^\circ C$. 
From (\ref{eqn_Gor'kov}), it is clear that for determining the Gor'kov potential, it is necessary to know the pressure field.

Acoustic levitation traps are regions in the field where the acoustic radiation forces converge. 
Consequently, objects placed in the levitation traps remain suspended in mid-air. Strong and stable acoustic traps can be created using optimization~\cite{marzo:nature}. 
Maximizing the converging radiation forces is equivalent to maximizing the Laplacian of the Gor'kov potential, given by:
\begin{equation} \label{eqn_lap}
\nabla^2 U= U_{xx}+U_{yy}+U_{zz},
\end{equation}
with the notation $U_a=\frac{\partial U}{\partial a}$ and $U_{aa}=\frac{\partial^2 U}{\partial a^2}$.

The pressure inside the traps tends to be very high, which can create disturbances for the levitating object. 
In order to avoid such disturbances, in addition to maximizing the Laplacian, following~\cite{marzo:nature}, we minimize the pressure as well.

The two arrays of transducers we employ in this study emit acoustic waves with constant amplitude and frequency. 
For simplicity, we assume the transducer to be a piston source, and use a model that neglects reflections and nonlinear effects, which is also in the interest of fast computation. 

The complex acoustic pressure of the $j^{th}$ transducer in the array can be written as~\cite{marzo:nature}
\begin{equation} \label{eqn_pressure}
p^j = e^{i\phi^j}M^j,
\end{equation}
where $\phi$ is the phase shift and M a complex number, specific to a transducer and a given point in space. 
Due to the rule of linearity of differentiation, $p^j_x = e^{i\phi^j}M^j_x$ also holds.
We calculate $M^j$ by the means of the example of Marzo et al.~\cite{marzo:nature}, assuming a circular piston source and using a single frequency far-field model. 
\begin{equation} \label{eqn_M}
M^j=P_0J_0(krsin\theta_j)\frac{1}{d_j}e^{ikd_j},
\end{equation}
where $P_0$ is a constant determined by the transducer power, $J_0$ is a zeroth-order Bessel function of the first kind, $k$ is the wave number, $r$ is the radius of the piston source, $\theta_j$ is the angle between the transducer normal and the focus point and $d_j$ the distance between the $j^{th}$ transducer and the focus point.

The total acoustic pressure field generated by $N$ transducers, assuming linear superposition of waves, is given by the sum of the pressures $p=\sum\limits_{j=1}^N p^j$, generated by individual transducers.  
As before, by linearity of differentiation, it holds that $p_x=\sum\limits_{j=1}^N p^j_x$. 
Going back to (\ref{eqn_lap}), it is clear that now, the Laplacian of the Gor'kov potential can be expressed as a function of only one variable - the phase shift: $\nabla^2 U= f(\phi_1,...,\phi_N)$. 
Hence, to produce a specific pattern in the acoustic field, the phase shift for each individual transducer needs to be calculated.

\subsubsection{Objective Function}
Numerical optimization methods make it possible to choose phase shifts for the individual transducers that best fulfill our predefined requirements.
To this end, following~\cite{marzo:nature}, we define a function which represents our problem objective - to minimize the pressure and maximize the Laplacian of the Gor'kov potential at a given point in space. 
This function is used as a criterion by the optimization procedure to select better rather than poorer solutions. 
To obtain a levitation trap at the point $\vec{q}$, we minimize the objective function
\begin{equation}\label{obj_fc}
O(\phi_1,...,\phi_N;\vec{q})=\mid p(\vec{q})\mid^2-\nabla^2 U(\vec{q}).
\end{equation}

Adding weights to control the relative strength of the trap in a particular direction and to balance the contributions of different terms, results in
\begin{multline}\label{obj_fc_weighted}
	O(\phi_1,...,\phi_N;\vec{q})=w_p\mid p(\vec{q})\mid^2\\-w_xU_{xx}(\vec{q})+w_yU_{yy}(\vec{q})+w_zU_{zz}(\vec{q}).
\end{multline}
We apply equal weights to each direction of propagation, hence generating a vortex trap.
An overview of different types of levitation traps is found in~\cite{marzo:nature}.

\subsubsection{BFGS Optimization}
The Broyden-Fletcher-Goldfarb-Shanno algorithm is an iterative method for solving nonlinear optimization problems. 
It belongs to the group of quasi-Newton methods, which have the advantage that the Hessian matrix is not evaluated at each step. 
Instead, an approximation generated by analyzing the successive gradient vectors is used, making the process more time-effective. 
This is favorable for solving our optimization problem, as we have to deal with a very large state space, consisting of 252 phase values. 
Thus, similarly to~\cite{marzo:nature}, we employ BFGS optimization to minimize (\ref{obj_fc_weighted}). 

\end{document}